\documentclass{article}
\usepackage[utf8]{inputenc}
\usepackage{url}
\usepackage[backend=biber,style=oscola]{biblatex}
\usepackage{xcolor}
\usepackage{graphicx}
\usepackage{enumerate}
\usepackage{array}
\usepackage{ragged2e}
\usepackage{amsmath}
\usepackage{appendix}
\usepackage{authblk}
\usepackage{mathtools}
\usepackage{mathptmx}
\usepackage{mathrsfs}
\usepackage{amssymb}
\usepackage{amsmath}
\usepackage{amssymb}
\usepackage{csquotes}
\title{Large Language Models as a (Bad) Security Norm in the Context of Regulation and Compliance\footnote{This is the submitted version, not the final version. That one will be published in the book \emph{Information Law and Policy Centre’s Annual Conference 2024: AI and Power: Regulating Risk and Rights}, published by University of London Press. As such, this one (the preprint) has fixes for mistakes, new additions, and it can be found in the book.}}

\author{Kaspar Rosager Ludvigsen\footnote{Assistant Professor in Medical Law, Durham Law School, Durham University (UK). PhD in Security and Law, Computer and Information Sciences, University of Strathclyde.}}

\date{2025}

\begin{document}

\maketitle

\begin{abstract}

The use of Large Language Models (LLM) by providers of cybersecurity and digital infrastructures of all kinds is an ongoing development. It is suggested and on an experimental basis used to write the code for the systems, and potentially fed with sensitive data or what would otherwise be considered trade secrets. Outside of these obvious points, this paper asks how AI can negatively affect cybersecurity and law when used for the design and deployment of security infrastructure by its developers.

Firstly, the paper discusses the use of LLMs in security, either directly or indirectly, and briefly tackles other types of AI. It then lists norms in cybersecurity, then a range of legal cybersecurity obligations from the European Union, to create a frame of reference. Secondly, the paper describes how LLMs may fail to fulfil both legal obligations and best practice in cybersecurity is given, and the paper ends with some economic and practical consequences for this development, with some notions of solutions as well.

The paper finds that using LLMs comes with many risks, many of which are against good security practice, and the legal obligations in security regulation. This is because of the inherent weaknesses of LLMs, most of which are mitigated if replaced with symbolic AI. Both also have issues fulfilling basic traceability obligations and practice. Solutions are secondary systems surrounding LLM based AI, fulfilment of security norms beyond legal requirements and simply not using such technology in certain situations.

\end{abstract}

\section{Introduction}

We start the paper with a quote from the cybersecurity consultant \emph{Colin Coghill}:

\emph{''how am I supposed to manage IT risk at my company when EVERY SINGLE VENDOR is throwing our private data into AI models that can be tricked into leaking it.''}\footnote{See \url{https://mastodon.nu/@mugginsm@cloudisland.nz/112453455981728016}, last accessed 16 December 2025.}

This central question is the core of this text; to focus on law and technology and their interaction, and how these AI models function in the context of the development of (cyber)security systems. 

Due to the state of the world at a supply chain and systems level, security is needed at every level, and for every system.\footnote{\cite{tan_advanced_2025}.} Adversaries always seek to use vulnerabilities, which, regardless of where in the system they are,\footnote{Unless it is in a sandbox or walled garden, terms which merely refer to sections which cannot reach other parts of any system.} lends themselves well to become holes which can be used to escalate an attack. This is the primary concept that motivates mandating either state of the art,\footnote{\cite{friedewald_conceptualising_2022}.} or best practice\footnote{\cite{niemimaa_information_2017}} security for every system possible, as the cost of the failure can be far higher than implementing it.

Large Language Models (LLM) are not a fully new phenomena, originally created for linguistics research from the notions of neural network models,\footnote{\cite{vaswani_attention_2023}.} which has become popular at large. A key indicator of their abilities lies in their structure, which as complicated systems require specific training and transformers, with commercial products like ChatGPT (any of them), and partially open source models like Llama, DeepSeek\footnote{See https://github.com/deepseek-ai/DeepSeek-V3, last accessed 31 January 2025.} and Qwen\footnote{See https://github.com/QwenLM/Qwen2.5, last accessed 31 January 2025.}  all possess. But if the training is poor, wrong, or missing, and if the transformer model is compromised or manipulated in any manner, then their efficiency and usefulness dramatically falls.\footnote{\cite{derner_security_2023}.} This is no surprise, as these models are built to statistically infer what to do next, be it in text, pictures, measurements, code, or otherwise, and this is a key debate in the academic discussion of artificial intelligence (AI), as none of this fulfils the idea of intelligence.\footnote{This is a contested area, but so far, they do not literally fulfil the requirements found in legislation, due them not being able to 1:1 occupy the role of humans, a physical and mental sense. They all need humans-in-the-loop, support systems, electricity and so on; things which means the are not anywhere near close.} Luckily, the EU has already defined AI in its AI Act, which we will get back to later.\footnote{REGULATION (EU) 2024/1689 OF THE EUROPEAN PARLIAMENT AND OF THE COUNCIL of 13 June 2024 laying down harmonised rules on artificial intelligence and amending Regulations (EC) No 300/2008, (EU) No 167/2013, (EU) No 168/2013, (EU) 2018/858, (EU) 2018/1139 and (EU) 2019/2144 and Directives 2014/90/EU, (EU) 2016/797 and (EU) 2020/1828 (Artificial Intelligence Act).}

In the security and coding space at large, major software service providers have started to offer AI support tools for coding, such as Copilot.\footnote{See https://github.com/features/copilot, last accessed 31 January 2025.} Copilot is an interesting example, since it is integrated into the world’s largest code repository, Github, and owned by Microsoft, the latter which is one of the world’s largest security providers, and the manufacturers of many types of infrastructure necessary to run environments to program in.\footnote{This can be in the form of their software for coding, their operating systems like Windows, and the fact that they also own Github, giving them a part of the entire supply chain of security in general.} Copilot, as one of the best case studies due to its ubiquity, has so far been found to be lacklustre, or even completely inefficient for anything but rudimentary and basic programming.\footcite{mastropaolo_robustness_2023, imai_is_2022, moradi_dakhel_github_2023} While these studies require further follow-up studies in many different environments to affirm their findings, this clearly indicates that LLMs are not fit to use when coding anything but the most simple of systems. We will return to this with some other arguments later.

Creating secure, and eventually safe systems requires rigorous principles, discipline and excellent project and risk management.\footcite{anderson_security_2020, anderson_certification_2009, anderson_economics_2006} This paper focusses on security, but safety is main motivation for the entire field – due to its potential. Security refers to the methods to keep systems from failing due to adversaries attacking them; conversely, safety refers to methods to keep systems from failing due to accidents, design flaws, or any other non-adversarial reason.\footcite{martinetti_redefining_2021, burton_mind_2020, leveson_safeware_1995, leveson_systems_2020}

Any system which has digital elements, that is software or hardware (and not necessarily a reliance on networks), can be attacked by adversaries, and designers, deployers, and users of these systems must prevent these digital failures from causing physical, financial, or reputational to anyone around or in the system.\footcite{lee_cyberphysical_2012}  

Safety engineering is the discipline which has had the most experience with building the necessary theory behind it to lower the rate of safety failures, but due to the security element present in all new technology, including LLM based AI, the acknowledgement of creating the best security possible to prevent safety failures is far more pressing than ever.\footcite{leveson_are_2020} The poor performance and rate of hallucination or worse actions made by these models speak for themselves, if they are ever to be deployed in surgical robots,\footcite{ludvigsen_dissecting_2020} nuclear reactors, planes, and other classic safety critical systems.

This paper contains a range of findings. Firstly, a discussion of what, which, and how norms matter in security. Its conclusions on how certain types of technologies can literally and directly violate the clear norms found, constitutes a major leap, and something which can lead to follow up empirical studies. Additionally, connecting these engineering-based norms to law should be considered another point of the text.
The paper is structured as follows:

Section 2 explains what LLM based AI versus other types of AI can be defined as. 

Section 3 delves into (cyber)security norms, defines and elaborates on what they are, and then explains how LLM fit into them currently, and what weaknesses they represent in a development environment. 

Section 4 discusses select security requirement found in EU Law.

Section 5 concludes the paper, with arguments for how LLMs fit into the current security climate, how these models and AI at large is topical due to choices made by both private and public actors, how companies should deal with the threat they face when using these models, and finally some thoughts on the costs for when these models cause security failures.

\section{AI Currently}

This sections paints a concise picture of what LLM based and other types of AI can be defined as.

To illustrate the idea, we use the definition found in Article 2(1) of the AI Act:\footnote{For relevant perspectives on the AI Act, see \cite{almada_regulation_2023, almada_delegating_2024, almada_eu_2022}.}

\emph{‘AI system’ means a machine-based system that is designed to operate with varying levels of autonomy and that may exhibit adaptiveness after deployment, and that, for explicit or implicit objectives, infers, from the input it receives, how to generate outputs such as predictions, content, recommendations, or decisions that can influence physical or virtual environments;}

As we will see below, it easily covers both categories of AI in this paper. More importantly, the definition is future-proof and, quite surprisingly, technology neutral.\footcite{almada_two_2024, koops_should_2006, ohm_argument_2010, reed_taking_2007}  

The latter refers to the notion that the types of AI available can change; what matters are  the abilities and perceived qualities these possess, which the wording above fully shows. If the analysed systems fulfil these, then they are included and must be regulated as AI. Additionally, the AI Act does not discern between whether these systems affect the physical world, such as robots, or the digital world, such as AI in videogames, or something in-between or which indirectly affects the world at large, such as automation in stock purchases\footcite{goldblum_adversarial_2021} or tax systems.\footcite{marr_youre_2012} All are covered.

\subsection{LLM based AI}

To preface the initial perspectives and descriptions of this type of AI, let us start with a quote that moves entirely in the other direction:

\emph{“Now the Web at large is full of slop generated by large language models, written by no one to communicate nothing.”}\footnote{Quote by Robyn Speer, see https://github.com/rspeer/wordfreq/blob/master/SUNSET.md, last accessed 31 January 2025.}

This quote resonates well with the dead internet theory,\footcite{walter_artificial_2024} and while not entirely relevant for LLMs in the context of security, it rings a necessary bell. LLM based AI does not need to purely be meaningless, created for no one, and to contribute nothing to society. But the risk of this happening in most of the areas which it has been deployed in is real, and should be mitigated, even if this fate is what the developers of it may intend, regardless of their statements. 

As mentioned in the introduction, LLM models require several parts which must all work in tandem to create the outputs, and in this case, these can be explained as \textit{the dataset}, sometimes called a \textit{foundational model} or by other terms, \textit{the transformer}, and the \textit{additional support systems} around it. If you run the first two, in some combination on your own hardware, you will not interface with the support systems. But if implemented into an app or used in a browser, then there are mechanisms on top of the actual model, as recently seen with DeepSeek.\footnote{Wang, Vivian, ’How Does DeepSeek’s A.I. Chatbot Navigate China’s Censors? Awkwardly.’, \url{https://www.nytimes.com/2025/01/29/world/asia/deepseek-china-censorship.html}, last accessed 31 January 2025. This is not limited to this model, all others who have been build into apps or similarly require online connections can face this issue.} 

These can materialise into changes to the output from the model, either for political, practical, or other censoring purposes – or simply to filter out certain types of meaningless outputs. From those that build and understand these models natively, in linguistics, we can view it as something that takes inputs, in the form of language (of any kind, be it pictures or text), interprets based on its dataset, makes it run through the transformer, and then brings it out as an output.\footcite{rosenfeld_whose_2024, birhane_large_2024, craft_language_2020} 

Obviously, it fits the criteria mentioned earlier, the AI Act was likely shaped around it in a way, but it tells us immensely much about the quality of said output. It can only be as good as the dataset and its transformer lets it be, and it cannot contain any element of plasticity,\footcite{leveson_safeware_1995} a human quality that is central to safety, which refers to immediate adaption, understanding, and reaction to something which malfunctions or otherwise does not work as intended. 

For this paper, we note that the dataset these models use contains code. The generic models provided by most companies will have a set amount, with specialised models like Copilot will container a far larger amount. As indicated in the introduction however, it does not so far seem like this specialisation provides larger efficiency, especially for code for fields like security. This comes down to the structure of LLMs, as they are large and still have to answer a vast amount of questions. Smaller specialised models, Small Language Models (SLM),\footcite{schick_its_2021, zhang_tinyllama_2024, zhu_llava-phi_2024} may be better suited for such tasks. In those situations, while they might only be able to answer questions about a specific language, or even just as specialised application (like security), they may be able to outpace the larger models without costing much to develop or deploy. Despite the obvious advantages of using SLM over LLM as wording, many smaller specialised models may opt for LLM, or even just AI, even though they are SLMs in reality.

LLMs are treated as oracles,the same way search engines used and still are, and come with the exact same drawbacks.\footcite{gonzalez-briones_legal_2022, hauglid_doctor_2023}

It allows us to make use of the research done on these oracle systems,\footcite{ferreira_exhausted_2022} even if the technology behind it is vastly different. The consequences of its usage when it is used to learn or otherwise find new sources of information therefore may be similar.

\subsection{Other types of AI}

If we look back at the definition of AI in Article 2(1) of the AI Act, we can see that this category encompasses every other type of system with the qualities brought forth. Any recommender system that makes use of simple machine-learning,\footcite{jaton_assessing_2021, lehr_playing_2017} more complicated neural-networks\footcite{ilyas_adversarial_2019} that are yet to be as complicated and massive as LLM based systems, and, most importantly, symbolic AI.\footcite{ilkou_symbolic_2020, feldstein_mapping_2024} 

The first concepts here afford the same risks and issues seen in LLMs, albeit they are simpler and far easier to control, while symbolic AI is on the completely opposite end of the spectrum. Symbolic AI is especially interesting here to discuss, because they represent fully predictable systems. 

Instead of relying on datasets to statistically give likely answers, as seen with the rest of the models here or LLMs, symbolic AI focusses on only providing answers within what they are programmed to do. We can see this as the split between deductive\footcite{lindahl_deduction_2004} and inductive reasoning\footcite{hunter_no_1998} – symbolic AI is fully deductive, even if fed with new information, it can only give answers within this narrow reasoning pattern. Conversely, all other models will mostly argue inductively, based on new inputs which then hermeneutically changes the next output, or sometimes potentially abductively,  a combination of both reasoning methods.

From the perspective of security, systems which can concretely advice developers with code are relevant – but they are not prevalent, and as seen with Copilot, perhaps not ready. More often than not, companies will have internal guides, “wikis”, and the like, not fully automated systems. Whether this is changing going forward is currently unknown.

\section{LLMs in Security}

Whether public or not, the usage of LLM based AI has increased in many parts of society, used to summarise or draft texts, transcribe audio, code and so forth.\footnote{With some success in fields like medical notation, see this initial study - \cite{asgari_framework_2024}.} This section gives a brief account of what we know currently when it comes to its usage in security. It then also dives into the vaguely defined area of security norms.

\subsection{Overview}

AI are, or can be used at various stages of development, deployment, and during adversarial attacks and failures. Each require its own small subsection, as it plays potentially different roles. These considerations are both concurrent and likely applicable in the future.

\subsubsection{Development}

AI may be used to write, co-write, assist, review, or otherwise attempt to improve the development process. The issue with almost all those steps, is whether the AI is able to truly assist or really accomplish anything, and whether this leads to either clear or hidden consequences in the software itself. LLM based AI may directly take existing code, and while it can run, it could contain known vulnerabilities, which must then be found at a later stage, or not, if the manufacturer does not properly test the system. It may then be found by a Notified Body in a European setting if part of various product regulatory schemes\footnote{This is seen in everything from medical device legislation, vehicles, and the machinery rules at large.} leading to either a denial of certification or a much lengthier process. If AI is used to choose the type of encryption, it may use outdated libraries or implement it poorly, and teams without competent cryptography specialists will cause the whole system to suffer before certification and potentially more during deployment. It may also cause the exact same result when co-writing, giving poor or subpar solutions to the questions asked, which requires more man-hours due to the need of senior developers to guide the users of LLM based AI.

\subsubsection{Deployment}

As of the time of writing, using LLM based AI in deployed security is only at a theoretical stage. The closest we get is existing ML based systems,\footcite{prokos_squint_2023} which could be replaced with similar LLM enabled systems. This will have the same downsides, require higher energy usage\footcite{samsi_words_2023} unless it is a specialised SLM, and may not be fit for purpose depending on what foundational models are used. 

Additionally, like above, legal rules still strongly impact how security must be deployed, which can directly limit the use of LLM based AI anywhere in the security supply chain. This could be in situations where the models actively create vulnerabilities, in real-time or otherwise, which would cause the developer to be liable for any third-party losses\footnote{Other responsibility regimes, product or otherwise, even without security specific provisions, would cause similar patterns of liability and potential damages.} due to their responsibility in both the Cyber Resilience Act (CRA),\footnote{REGULATION (EU) 2024/2847 OF THE EUROPEAN PARLIAMENT AND OF THE COUNCIL of 23 October 2024 on horizontal cybersecurity requirements for products with digital elements and amending Regulations (EU) No 168/2013 and (EU) No 2019/1020 and Directive (EU) 2020/1828 (Cyber Resilience Act).} the NIS2 Directive (NIS2),\footnote{DIRECTIVE (EU) 2022/2555 OF THE EUROPEAN PARLIAMENT AND OF THE COUNCIL of 14 December 2022 on measures for a high common level of cybersecurity across the Union, amending Regulation (EU) No 910/2014 and Directive (EU) 2018/1972, and repealing Directive (EU) 2016/1148 (NIS 2 Directive).} and specialised legislation like the Medical Device Regulation (MDR).\footnote{REGULATION (EU) 2017/745 OF THE EUROPEAN PARLIAMENT AND OF THE COUNCIL of 5 April 2017 
on medical devices, amending Directive 2001/83/EC, Regulation (EC) No 178/2002 and Regulation (EC) No 1223/2009 and repealing Council Directives 90/385/EEC and 93/42/EEC, L 117/1.}

If we move beyond what we can currently do, and think about future implications, then we could start by looking at the generative aspects in a security deployment context.
Attempting to predict what files, connections, requests and so on that would be necessary for a set of parameters, does not seem to be what any of the current AI are capable of, whether in real-time or otherwise. The other issue for that usage is calibration and verification,\footcite{brown_infrastructure_2020, v_gayetri_devi_efficient_2018} as all these parameters need to be controlled and understood, something which LLM based AI is currently not necessarily capable of providing. The exception to this would be evaluation AI of these systems, but this still requires that there is no Black Box element anywhere in the supply chain, at least after the last evaluation by such as secondary AI system. 

\subsection{Weaknesses of LLMs}

LLMs suffer from all the same disadvantages that machine-learning and neural-network\footcite{real_physical_2021, elsayed_adversarial_2019} based AI do. This includes data poisoning, which for LLMs can include input data, the actual data it was based on when it was trained, or even the foundational model it can be partially based on. The latter is a very important point, as any information about the security that is being developed, parameters of the system around it or which it gives security to, or worse, \textit{trade-secrets}, is possible to cheat out of any LLM. 

\subsection{LLMs in Security Norms}

We focus on norms in this section.\footnote{For some examples, see \cite{finnemore_constructing_2016, kouloufakos_untangling_2023}.} It is worth keeping in mind that, depending how one views legal methods and jurisprudence at large, norms are either actual law, or have nothing to do with law at all.\footnote{Norms are defined and understood very different in Legal Realism compared to Positivism, whether traditional or otherwise, see \cite{hart_scandinavian_1959}. See also \cite{pihlajamaki_against_2004, alexander_comparing_2002}.}  For this paper, we focus on how engineers and developers view norms necessary to develop systems for security, and these are as much closer to what can be defined as “best practice”, though not entirely. The latter is often defined by lobbyists or technical specialists, and this leaves out diverging views, whereas norms here refer to broader concepts and ideas. 

\subsubsection{Security Norms}

Creating the best security possible requires vigilance through the development, deployment, and sunsetting process.\footnote{For a list and analysis of norms at large, see \cite{madnick_evolution_2024}.} Norms in security are not well studied, though they are covered well without using the term in technical papers and meta studies, and in risk management settings. The listed norms below are not final, nor are all possible norms listed, and they will not always apply cumulatively.\footnote{Additional research on the norms and their role as such is seen elsewhere in Human Computer Interaction studies, though more research on the interface between this and law is needed.} 

As such, there are individual norms for each of these steps which must be followed, of which we cannot realistically list all. But there are also some overarching norms which should be followed that apply to all aspects of security, namely:

•	\textit{1. To not rely on secrecy for core parts of the system.}\footnote{Also called \emph{Kerchkoff’s Principle}, see \cite{kerckhoffs_cryptographie_1883, anderson_open_2005}.} This amounts to using well-tested and known cryptographic schemes over relying on secrecy, as strength which is thoroughly tested and even after that is still hard to break, is to be preferred over secret but potentially easily breachable schemes. This can be extended to structure, deployment type and so on; what matters is to not rest on one’s laurels. 

•	\textit{2. To follow best practice (if possible).}\footnote{This term is wide and defined in a couple of ways, see \cite{niemimaa_information_2017, mccarthy_summary_2014, saint-germain_information_2005}.}  Well-proven approaches at any point are preferable, but it is a process – reaching it is difficult. This can be both organisationally or in the choice of technology, such as encryption,\footcite{shurson_european_2024} access control, choice of hardware and so on.

•	\textit{3. To ideally follow state of the art.}\footcite{friedewald_conceptualising_2022} This specifically refers to which technical approaches to take, not organisational. This is usually in the context of what type of defences which should be used, recently seen in 2-factor authentication.

•	\textit{4. To understand every single component, action, possible weakness and role of the system.} A matter of control, this norm is implicit, and it also ties security together with digital forensics\footcite{losavio_internet_2018} and auditing.\footcite{ryoo_cloud_2014} If there is no control and perception of what is being created, any system will experience difficulty passing auditing (of any kind) and will give investigators a headache when the system is examined after a failure.

•	\textit{5. To be available when necessary.} Offering security often includes servicing agreements, throughout all part of the system’s lifecycle; this entails such a norm, and it is more implicit than anything else. Not adhering to it can be costly.

These are all very idealistic, and this is by design, as security engineering always strives towards the best possible state. 

We should also mention norms within the four stages above.

•	Development. During the design of security, a primary norm is to include security as early as possible when designing any system.\footcite{anderson_security_2020} Not only will the costs exponentially increase for whoever wants security, it will be poorer the longer they wait. Another key norm is understanding dependencies, and their relationship with both the surrounding system, and any nearby critical infrastructure. Not only would it be costly for developers to harm society at large if their system is a gateway to damage, for example, a hospital or a national ISP. This must be understood and prevent by all means possible and can be seen as a part of a greater control and information norm, which developers must always possess. 

•	Deployment. A key norm is that deployment must be secure and safe, with the latter indicating that security failures should not cause safety failures. This refers to situations where the deployment of a system causes damage, be it financial, physical or otherwise. This must be prevented in real-time. Secondly, the security, which is deployed, should not interfere with other systems, which can be seen as an expansion of the norm above. 

•	Sunsetting. There are far fewer norms in this area, but norms that apply to legacy systems function here, such as preventing network access for legacy systems, publicly stating and explaining the existing vulnerabilities, and responsibly ending the life of systems and replacing them without causing gaps that can be used by adversaries.\footcite{ohm_legacy_2022}  

We can now ask: How do these selected norms fit with LLM based AI?

\subsubsection{The Case of LLMs and AI in general}

As we saw earlier, LLMs give answers based on their design and their foundation. This already leaves much to be desired. None of the norms above can include AI that makes use of LLMs, as they are prone to give wrongful or hallucinated answers. If symbolic AI is used however, it is a different situation, as they can be used as warning systems which can alert personnel to solve problems in real-time. Such AI can also be used for risk management when it comes to surveillance of users, activities in system and so on. The latter is also possible with tightly controlled machine-learning and neural-network based AI but is not the case when involving LLM.\footnote{It may be possible if the model is an SLM, is heavily supervised by support systems which may be symbolic AI or simpler software themselves, and perhaps with additional supervision from humans. But at this point it may be cheaper to not use them at all.} 

LLM based AI violate \emph{Kerchkoffs Principle} unless made in-house and fully understood. Good security mandates no Black Boxes, and these models are mostly still Black Box when it comes to the data used, and some of the structure in the transformer, though as mentioned in the introduction, some are more open about this than others.

LLM based AI violate deployment placed norms, simply because they have such a high risk to lie deliver wrongful answers. Anything beyond a couple of percent is unacceptable in any environment with queries that go beyond thousands per second, and for security this is no different. An error rate near 99 percent is still not enough for the often-critical environments that security works in.

\section{Legal Obligations in Security}

Not all jurisdictions have made security specific legislation; it may be instead placed elsewhere, such as under national traditional security, surveillance laws, or similar, but EU Law has in the last couple years made remarkable progress towards the opposite.\footcite{ludvigsen_creating_2024} We will look at some examples in the context of LLM based AI and otherwise.

\subsection{Statutory Law}

The CRA (Cyber Resilience Act) regulates cybersecurity at large, and as such sets the tone and makes minimal levels of security required in products.\footcite{chiara_understanding_2025, colonna_end_2025, eckhardt_eus_2023}  It does not regulate the infrastructure of public or private bodies’ own internal security, though it might do so indirectly. A set of requirements which align with the norms we discuss above can be found in Annex I – not complying with these is part of the penalty system in Article 64. These can therefore be seen as hard minimum requirements for all products and devices with digital elements. Annex I has two elements, Part I and Part II, which both must be complied with. The first covers design and deployment, and all follow the general norms given in last section. Worth noting is that all elements require precision, and as we know, LLM based AI is not capable of this currently, meaning that they cannot replace existing systems when complying with Annex I. Additionally, Annex I, Part II discusses the handling of vulnerabilities, which connects to the fine balance from our state of the art norm earlier. These are practical decisions, or they may be economical, and none of these can again be taken by AI. 

The NIS2 Directive, while needing to be implemented in member states, has in its Article 21 set out a similar approach to the necessary “Cybersecurity risk-management measures”. In Article 21(1), second part, state of the art as the goal and primary target is laid out. Article 21(2) contains minimum requirements of the risk management measures necessary, which all follow existing best practice notions, while Article 21(4) is a strong compliance measure. If Article 21(2) is not followed, then member states can force all those covered to do so by all means necessary. To this, it is worth mentioning that what NIS2 is aimed at, is defined in Annex I.\footnote{These categories are broadly energy, transport, banking, financial market infrastructures, health, drinking water, waste water, digital infrastructure, ICT services management, public administration, and space. For different perspectives in an adjacent area, see \cite{kun_challenges_2024}.} As the covered group is large and includes all of society at large, security providers will naturally, be it directly or indirectly, be included. For the indirect part, consider that security providers will need comply with the NIS2 requirements even if they only contractually deliver services to those covered by Annex I, and should therefore prepare for this situation. 

The AI Act also has a specific cybersecurity provision in Article 15. Article 15(1) uses the idea of appropriate cybersecurity, which is weaker than state of the art; a surprising choice which does not align with the CRA or NIS2. Conversely, the Article does focus on resilience, which is the notion that systems can fail gracefully and otherwise partially failed in a controlled manner.\footnote{See two different perspectives, \cite{toftegaard_operational_2024, bygrave_cyber_2022}.} The AI Act only mandates most of these elements on High Risk AI, which matters less when we now know that they are by themselves covered by other cybersecurity regulation in the EU. Article 15(5), part 3, is worth focussing on because it mandates that all types of unique attacks on machine-learning,  neural-network or LLM based AI should be prevented where appropriate. These amount to technology specific instructions that align closely with the best practice norm. How this is complied with going forward is unknown, as several of these are hard problems to solve in a technical context.

\subsection{Contracts}

Security is mostly provided by other parties than those that need it, and as such contracts play a central role on top of statutory rules, and the norms that exist within the field. 
Contrary to the norms, contracts can stipulate poorer security than state of the art or even best practice. This can only be done when not covered by other rules, meaning in an EU context, it must not be part of the parties in Annex I in NIS2, and not be security in the products covered by the CRA. Regarding LLM based AI, this category is rather limited, but theoretically possible.
Contracts can add additional compliance measures in the form of stipulated damage coverage, clear liability for poor security, and even add technological requirements that would not otherwise be there. If a contract for example stipulates state of the art, then this norm, if covered by the CRA, will exist in three different ways. Firstly, contractual, then legal and then through security norms, creating an incitement and compliance structure which internal security providers in companies or public administration will not have, and which is far stronger due to its three layers.
If we look back on the quote that started this paper, contracts themselves do not protect those that pay for security to not leak their internal information. For this you need statutory law, as not all jurisdictions protect well against theft or leaking of trade-secrets or similar information, and data protection is not covered everywhere either. Contracts that involve larger supply chains may need to require this to not compromise the chain at large, and to pre-emptively fulfil the legislation which may exist and one or both jurisdictions which they start and end in respectively.\footcite{ludvigsen_preventing_2022}  

\subsection{Losses}

Failure to comply with the law can result in penalties, and at worst, forceful withdrawal or bans of products. Losses incurred by successful adversarial attacks, which can be proven to be caused by violation of security norms, law, or contractual stipulation, can lead to great financial losses or bankruptcy. Using LLM based AI to write code, for real-time surveillance or otherwise, risks concrete losses due to their weaknesses and known inability to fulfil basic security norms.
The risk of using AI for any critical security role is not new, as automation in security has been a topic for as long as it has existed. Due to the larger datasets available, and the willingness to deploy these systems at scale, most types of AI will see greater usage going forward. There may be incentives to even use LLM based AI for the tasks which they are poor at, if their perceived efficiency outweighs the potential losses.
This mimics other societal trends with increased digitalisation, but goes against security norms, depending on the develop of current and future AI.

\section{Conclusion}

This paper has shown that LLM based AI cannot fulfil the same cybersecurity norms as other types of AI can, due to their irregular answering patterns, black box and uncontrollable nature, and these qualities cannot fulfil legal rules in both the CRA, NIS2 and the AI Act in EU law. Additionally, LLM based AI will make development, deployment, and dismantling of security systems harder due to these exact problems. The paper also laid out a range of security norms which neatly connects to the obligations mentioned, which are to not rely on secrecy for good security, to follow best practice, to follow state of the art, to strive to always fully understand the entire security system, and to be available. The paper finally also showed some additional norms that only occur during the three aforementioned parts of the lifecycle of a security system, and how these can also clash with the surrounding technical reality of LLM based AI.

\subsection{Political Choice}

Lastly, a final concluding comment.\footnote{This section does not appear in the final version.} Deploying or even allowing LLM based AI is a political choice. They are based on copyright protected data, without a doubt, and this would normally equate a necessary contractual relationship between the rightsholders and the AI developers. But this has yet to materialise for all the material they make use of, and continuing their usage and deployment is therefore also a continuation of violation of intellectual property protection. Perhaps even an aspect of enshittification.  In a security context, this is less of an issue, as sharing code, especially in smaller amounts, or libraries, or organisation techniques is all done freely and widely, and is currently a central cultural aspect of cybersecurity at large. There is therefore less of a clash between this issue and security as such, which contrasts the rest of this paper.

\newpage

\printbibliography
\end{document}